\documentclass[fleqn,twoside]{article}
\usepackage[headings]{espcrc2}
\usepackage{amsmath}
\usepackage{subfigure}

\def\lr{\left( }
\def\rr{\right) }

\def\beq{\begin{equation}}
\def\eeq{\end{equation}}
\def\bea{\begin{eqnarray}}
\def\eea{\end{eqnarray}}

\usepackage{graphicx}
\usepackage[figuresright]{rotating}

\title{Perspectives for inclusive quarkonium production in photon-photon
 collisions at the LHC}

\author{M.\ Klasen\address{Laboratoire de Physique Subatomique et de Cosmologie,
 Universit\'e Joseph Fourier / CNRS-IN2P3 / INPG, 53 Avenue des Martyrs, F-38026
 Grenoble, France}\thanks{Research supported by the French Ministry for Higher
 Education and Research, CNRS-IN2P3 and ANR.},
        J.P.\ Lansberg\address{Institut f\"ur Theoretische Physik, Universit\"at
 Heidelberg, Philosophenweg 19, D-69120 Heidelberg, Germany}\thanks{Talk given by
 J.P.\ Lansberg at the Workshop on ``High-energy photon collisions at the LHC'',
 April 22-25, 2008, CERN, Switzerland.}}      
\runtitle{Perspectives for inclusive quarkonium production in photon-photon
 collisions at the LHC}
\runauthor{M.\ Klasen and J.P.\ Lansberg}

\begin{document}

\begin{abstract}
We report on the current status of knowledge on inclusive quarkonium production
in high-energy photon-photon collisions. As a perspective for the LHC, we compute
various production cross sections via direct photon-photon fusion in
ultra-peripheral pp, pA and AA collisions at the LHC using the tree-level
quarkonium amplitude generator {\tt MadOnia}.
\vspace{1pc}
\end{abstract}

\maketitle

\vspace*{-100mm}
\noindent LPSC 08-068
\vspace*{80mm}

\section{Introduction}

In a few months the LHC will start operating and provide physicists with a
tremendous amount of new experimental information on the fundamental particles
and their interactions. In this context, much is expected from the LHC to
elucidate the still ill-understood quarkonium production mechanism (see
\cite{Lansberg:2006dh,Brambilla:2004wf,Kramer:2001hh} for recent reviews). This
is of particular importance in view of the possible suppression of quarkonium
production after quark-gluon plasma formation in heavy-ion collisions. One of the
questions which remains to be answered is in particular the relevance of
color-octet transitions in different production regimes or equivalently of the
${\cal O}(v)$ corrections in Non-Relativistic Quantum Chromodynamics (NRQCD)
\cite{Bodwin:1994jh,Nayak:2006fm}.

The large amount of quarkonia that will be produced in the pp, pA and AA
modes at the LHC will allow for even more precise analyses than those performed
recently at Run II of the Tevatron \cite{Acosta:2004yw,Abazov:2005yc}. We expect
cross section and polarization measurements of inclusive, prompt and (hopefully)
direct yields of the $J/\psi$, $\psi'$, $\Upsilon$, $\Upsilon'$ and $\Upsilon''$
mesons. In addition, analyses of their associated production with a heavy-quark
pair \cite{Artoisenet:2007xi,Artoisenet:2008tc} or of other channels would
certainly be as interesting and in some cases better suited to disentangle the
color-singlet and color-octet transitions and to assess the importance of
$s$-channel cuts \cite{Haberzettl:2007kj,Lansberg:2005pc} and of the rescattering
with comovers \cite{Hoyer:1998ha,Marchal:2000wd}.
\setcounter{footnote}{0}
Recently, several quarkonium production cross sections have been evaluated at
next-to-leading order (NLO) of perturbative QCD \cite{Klasen:2004az,%
Klasen:2004tz,Campbell:2007ws}, complementing the pioneering
work for color-singlet production in direct $\gamma p$ collisions
\cite{Kramer:1995nb}. The common feature of these calculations is the significant
size of the NLO corrections, in particular for large transverse momenta of the
quarkonia. The color-singlet prediction can thus be brought considerably closer to
the data. As of today, only the color-octet contributions to direct
$\gamma\gamma$ collisions have been evaluated at NLO \cite{Klasen:2004az,%
Klasen:2004tz}.\footnote{Note, however, the recent preprint \cite{Gong:2008ft}.}

Very little is known yet about NLO effects on the polarization of the produced
quarkonia. However, the leading order (LO) NRQCD prediction including sizable
color-octet contributions is known to be in conflict with the measurements e.g.\
in hadroproduction \cite{Abulencia:2007us,Acosta:2001gv}. A recent NLO evaluation
for $J/\psi$ mesons produced directly as color-singlet states in gluon-gluon
fusion changes the LO prediction from almost purely transverse to rather
longitudinal polarization at NLO \cite{Gong:2008sn}. However, the corresponding
unpolarized differential cross section still falls short of the data, preventing
any conclusive interpretation of this result \cite{Campbell:2007ws}. A forthcoming
study of both the differential cross section and the polarization at NLO
complemented by dominant NNLO corrections might help to elucidate the situation
further, at least for hadroproduction \cite{Artoisenet:2008fc}.

In the absence of an International Linear Collider (ILC), the LHC will, at least
in the short term, present an almost unique environment for high-energy
photon-photon collisions \cite{Baltz:2007kq}. These tend to produce much cleaner
events than those produced in inelastic hadron collisions. They may also help to
validate the only experimental analysis of quarkonium production in photon-photon
collisions to date \cite{Abdallah:2003du}, which clearly required important
color-octet contributions \cite{Klasen:2001cu}.
We therefore focus here on quarkonium production in photon-photon collisions at
the LHC. First, we recall the theoretical status along with more specific results
related to the LEP II analysis. We then present original results for
ultra-peripheral $pp$, $pA$ and $AA$ collisions obtained with the tree-level
quarkonium amplitude generator MadOnia \cite{Artoisenet:2007qm}, before we
present our conclusions and an outlook.

\section{Inclusive quarkonium production in photon-photon collisions}

Photons can interact either directly with the quarks participating in the
hard-scattering process (direct photoproduction) or via their quark and gluon
content (resolved photoproduction). Thus, inclusive quarkonium production in
photon-photon collisions receives contributions from the direct, single-resolved,
and double-resolved channels. All three contributions are formally of the same
order in the perturbative expansion and must be included. The $J/\psi$ mesons can
be produced directly, or via radiative or hadronic decays of heavier charmonia,
such as $\chi_{cJ}$ and $\psi^\prime$ mesons, or via weak decays of $B$ hadrons,
where the latter can often be safely neglected. The cross sections of the four
residual indirect production channels may be approximated by multiplying the
direct-production cross sections of the respective intermediate charmonia with
their decay branching fractions to $J/\psi$ mesons.

Invoking the Equivalent Photon Approximation and the factorization theorems
of the QCD parton model and NRQCD, the differential cross section can be written
as
\begin{eqnarray}
\lefteqn{d\sigma(AB\to AB+H+X)=\int dx_Af_{\gamma/A}(x_A)}
\nonumber\\
&&{}\times\int dx_B
f_{\gamma/B}(x_B)
\sum_{a,b,d}\int dx_af_{a/\gamma}(x_a,M)
\nonumber\\
&&{}\times\int dx_bf_{b/\gamma}(x_b,M)
\sum_n\langle{\cal O}^H[n]\rangle\nonumber\\
&&{}\times
d\sigma(ab\to c\overline{c}[n]+d),
\label{e1}
\end{eqnarray}
where $f_{\gamma/A,B}(x_{A,B})$ is the equivalent number of transverse photons
radiated by the initial-state particles $A$ and $B$,
$f_{a,b/\gamma}(x_{a,b},M)$ are the parton densities of the photon,
$\langle{\cal O}^H[n]\rangle$ are the operator matrix elements of the $H$ meson,
$d\sigma(ab\to c\overline{c}[n]+d)$ are the differential partonic cross 
sections,
the integrals are over the longitudinal-momentum fractions of the emitted
particles w.r.t.\ the emitting ones, and it is summed over
$a,b=\gamma,g,q,\overline{q}$ and $d=g,q,\overline{q}$, with $q=u,d,s$.
To leading order in $v$, we need to include the $c\overline{c}$ Fock states
$n={}^3\!S_1^{(1)},{}^1\!S_0^{(8)},{}^3\!S_1^{(8)},{}^3\!P_J^{(8)}$ if
$H=J/\psi,\psi^\prime$ and $n={}^3\!P_J^{(1)},{}^3\!S_1^{(8)}$ if
$H=\chi_{cJ}$, where $J=0,1,2$.
With the definition $f_{\gamma/\gamma}(x_\gamma,M)=\delta(1-x_\gamma)$,
Eq.~(\ref{e1}) accommodates the direct, single-resolved, and double-resolved 
channels.
The presence of parton $d$ is to ensure that $P_T$ can take finite values.

\begin{figure}[h]
 \includegraphics[width=\columnwidth]{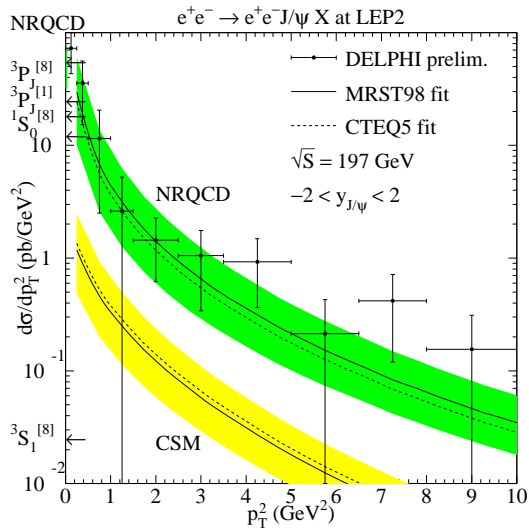}
 \caption{\label{fig:1}
 The cross section $d\sigma/dP_T^2$ of $e^+e^-\to e^+e^-+J/\psi+X$
 measured by DELPHI \protect\cite{Abdallah:2003du} as a function of $P_T^2$ is compared
 with the theoretical predictions of NRQCD and the CSM.
 The solid and dashed lines represent the central predictions obtained with the
 ME sets referring to the MRST98LO  (default) and CTEQ5L
 PDFs, respectively, while the shaded bands indicate the
 theoretical uncertainties on the default predictions.}
\end{figure}

In Fig.\ \ref{fig:1}, the $P_T^2$ distribution of
%
%%% Begin  Figure 1 %%%
%%% End of Figure 1 %%%
%
$e^+e^-\to e^+e^-+J/\psi+X$ measured by DELPHI \cite{Abdallah:2003du} is compared
with LO NRQCD and color-singlet predictions \cite{Klasen:2001cu}. The solid lines
and shaded bands represent the central results with MRST98LO parton densities and
their uncertainties, respectively, while the dashed lines represent the results
obtained with CTEQ5L parton densities. It is clear that at LO the DELPHI data
favor the NRQCD prediction, while they significantly overshoot the CSM one.
Note, however, that the associated production of $J/\psi$ mesons
with additional $c\bar{c}$ pairs can be numerically important
\cite{Qiao:2003ba}.

At NLO, the inclusive production of $J/\psi$ mesons with finite values of $P_T$
has so far only been studied in the color-singlet channel in direct photon-proton
\cite{Kramer:1995nb} and proton-proton \cite{Artoisenet:2007xi,Campbell:2007ws}
collisions and in
the color-octet channel only in direct photon-photon collisions
\cite{Klasen:2004az,Klasen:2004tz}. The technical difficulties that need to be
tackled in these calculations include the treatment of UV, collinear, soft, and
Coulomb singularities, the NRQCD operator renormalization, the mass factorization,
and the analytic evaluation of five-point one-loop integrals with UV, soft, and
Coulomb singularities.
As for the real corrections, the phase-space slicing method can be employed to
demarcate the regions of phase space containing soft and collinear
singularities from the hard regions, where the phase-space integrations can be
carried out numerically. One must, of course, verify that the combined result is,
to very good approximation, independent of the choices of the cut-off parameters
over an extended range of values.
One usually works in dimensional regularization in connection with the
$\overline{\rm MS}$ renormalization and factorization schemes, so that the NLO
result depends on the QCD and NRQCD renormalization scales $\mu$ and
$\lambda$, respectively, and on the factorization scale $M$ connected with the
collinear splitting of the incoming photons into massless $q\overline{q}$
pairs.
While the $\mu$ and $\lambda$ dependences are formally canceled up to terms
beyond NLO within direct photoproduction, the $M$ dependence is only
compensated by the LO cross section of single-resolved photoproduction.
By the same token, the strong $M$ dependence of the latter is considerably
reduced.
This is of crucial phenomenological importance, as may be appreciated by
observing that, at LO, the overwhelming bulk of the cross section of prompt
$J/\psi$ production in two-photon collisions is due to single-resolved
photoproduction.
The naive $K$ factor of direct photoproduction, defined as the NLO to LO
ratio, turns out to be very substantial at large values of $P_T$ because
fragmentation-prone channels start to open up at NLO.
In fact, at $P_T\gg2m$, the $P_T$ distribution of direct photoproduction at
NLO is rather similar to the LO result of single-resolved photoproduction both
in shape and normalization.

\section{Inclusive quarkonium production in ultra-peripheral collisions at the
 LHC}

We now turn to the calculation of quarkonium production through photon-photon
collisions in ultra-peripheral collisions of protons and heavy ions at the LHC.
We focus on the production of quarkonia with direct photons, as only these
processes have so far been implemented in the tree-level
quarkonium amplitude generator {\tt MadOnia} \cite{Artoisenet:2007qm}.
This Monte Carlo generator uses a simple method to automatically evaluate
arbitrary tree-level amplitudes involving the production or decay of a heavy
quark pair $Q\bar{Q}$ in a generic $^{2S+1}L_J^{[1,8]}$ state, i.e.\ the short
distance coefficients appearing in the NRQCD factorization formalism. The
approach is based on extracting the relevant contributions from the open heavy
quark-antiquark amplitudes through an expansion with respect to the
quark-antiquark relative momentum and the application of suitable color and spin
projectors. Several checks are performed to validate the calculation,
including those of gauge invariance and numerical cancellation of contributions
bound to vanish due to symmetry conditions. As a typical set for our numerical
input, we use a $J/\psi$-meson mass of $m_{J/\psi}=2m_c=3$ GeV,
electromagnetic and strong coupling constants of $\alpha=1/137$ and
$\alpha_s(\mu_r)$ with $\mu_r=\sqrt{m_\psi^2+P_T^2}$,
and color-singlet and color-octet operator expectation values of
$\langle O \left( ^3S_1^{[1]}\right) \rangle = 1.16$ GeV$^3$ and
$\langle O \left( ^3S_1^{[8]}\right) \rangle = 1.06\cdot10^{-2}$ GeV$^3$.

\subsection{Photon flux}

A reasonable analytic approximation for the photon flux in ultra-peripheral
collisions of protons and/or heavy ions $A$ and $B$ is obtained by integrating
the differential flux in impact parameter space over radii larger than $R_A+R_B$,
where the proton radius is $R_{\rm p}=0.6$ fm and the radii of heavier nuclei are
$R_{A,B}\simeq R_0 (A,B)^{1/3}$ with $R_0=1.2$ fm and $A,B$ the nucleon numbers.
The integrated flux is then given by
\bea
 f_{\gamma/A}(x)&=&\frac{2 Z^2 \alpha}{\pi x}
 \Big[\omega K_0(\omega) K_1(\omega) \\ && \hspace*{9mm} - \frac{\omega^2}{2}
 \lr K_1^2(\omega)-K_0^2(\omega)\rr\Big]. \nonumber
\eea
Here $Z$ is the proton or nuclear charge,
$x=k/E$ is the fraction of the beam energy $E$ of the protons or heavy ions
with mass $m$ carried by the almost real photon with energy $k$ and virtuality
$-q^2<1/R_{A,B}^2$, $K_{(0,1)}$ are the $0^{\rm th}$- and $1^{\rm st}$-order
modified Bessel functions of the third kind, and $\omega=xm(R_A+R_B)$.
The beam energies and corresponding expected luminosities for the LHC can be
found in Tab.\ 1 of \cite{Baltz:2007kq}. Note, however, that for protons we do
not use the spectrum above but rather the one in Eq.\ (D.7) of
\cite{Budnev:1974de}, both for pp and pPb collisions. In the latter case, this
gives us a first reasonable approximation of the photon flux,
but the difference in the minimum value of the impact parameter ($2 R_{\rm p}$
vs.\ $R_{\rm p}+R_{\rm Pb}$) should be taken into account in the future.

\subsection{Numerical results}

We present here our numerical results for $J/\psi$-production in direct
photon-photon collisions for three different hadron-beam combinations:
proton-proton (pp) collisions at a center-of-mass energy of $\sqrt{s}=14$ TeV,
proton-lead (pPb) collisions at a nucleon-nucleon center-of-mass energy of
$\sqrt{s}=8.8$ TeV, and lead-lead (PbPb) collisions at a nucleon-nucleon
center-of-mass energy of $\sqrt{s}=5.5$ TeV. The expected luminosities listed
in Tab.\ 1 of \cite{Baltz:2007kq} are $10^7$ mb$^{-1}$ s$^{-1}$, 420 mb$^{-1}$
s$^{-1}$ and 0.42 mb$^{-1}$ s$^{-1}$.

In order for the transverse momentum of the produced $J/\psi$ mesons to be
balanced, it must be produced in association with at least one other particle.
The direct production of a color-singlet with one or two gluons is forbidden
by color conservation and charge conjugation. The same holds for light $q\bar{q}$
pairs, which originate from gluon splitting. The dominant channels are thus
the associated production of a color singlet with a photon or an additional
$c\bar{c}$ pair or the associated production of a color octet with a gluon.
Note that in principle these final states are experimentally distinguishable.

The transverse-momentum spectra for $J/\psi$ mesons produced in pp, pPb, and PbPb
collisions are shown in Fig.\ 2 (top, center, and bottom) for central rapidity
%
%%% Begin  Figure 2 %%%
\begin{figure}
 \includegraphics[width=\columnwidth]{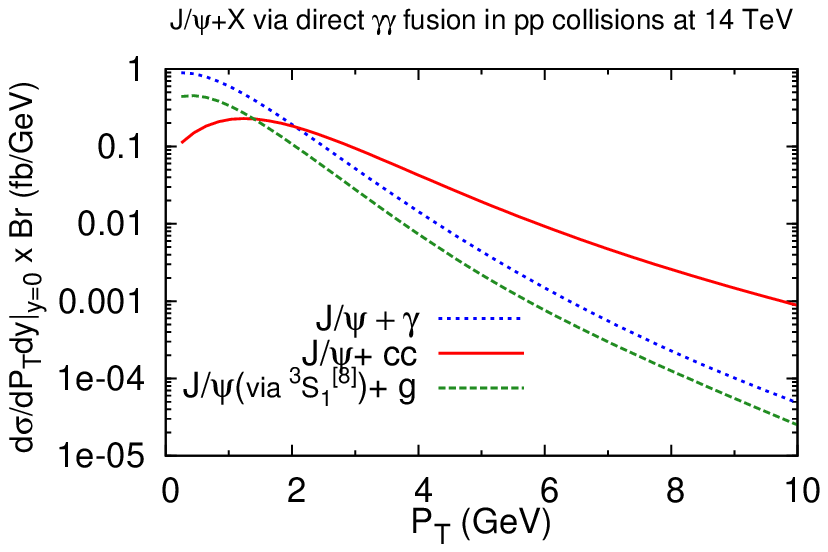}
 \includegraphics[width=\columnwidth]{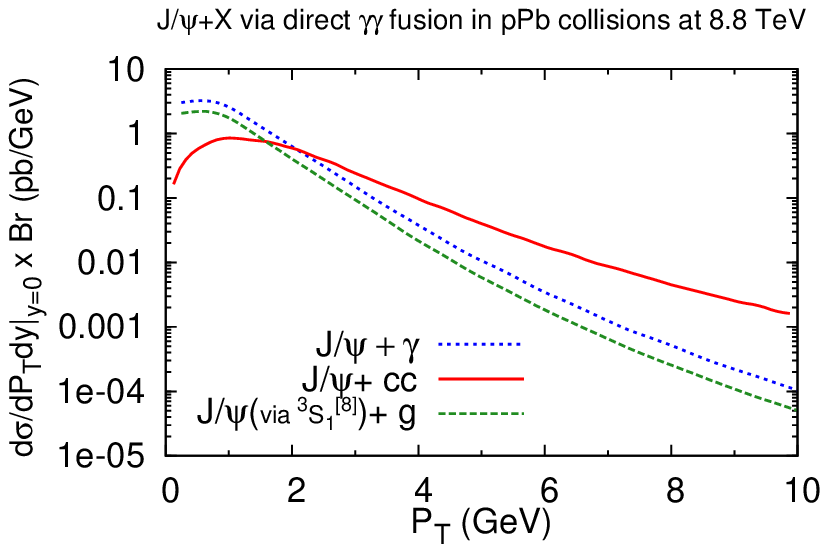}
 \includegraphics[width=\columnwidth]{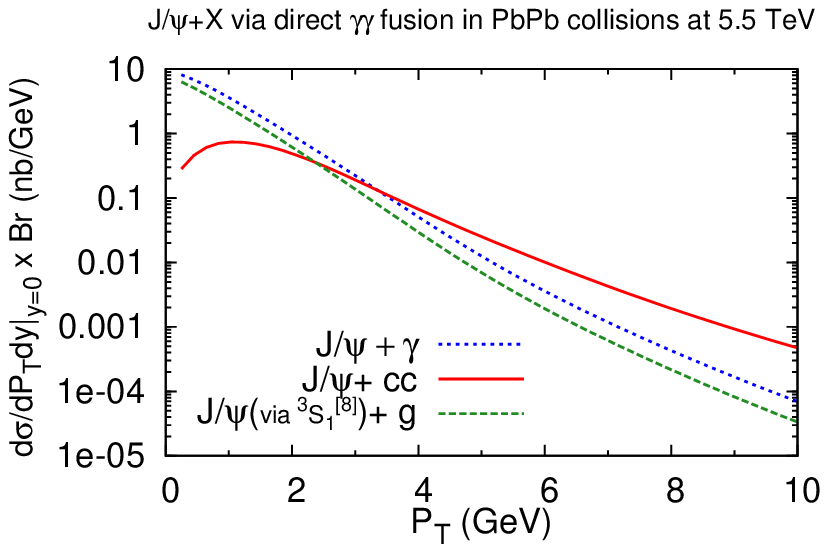}
 \caption{\label{fig:2}Transverse-momentum spectra of $J/\psi$-mesons produced
 in association with photons (dotted), $c\bar{c}$ pairs (full) and gluons (dashed)
 in pp (top), pPb (center) and PbPb (bottom) collisions. While the first two
 processes produce directly a color-singlet, the latter produces necessarily an
 intermediate color-octet state due to color conservation.
}
\end{figure}
%%% End of Figure 2 %%%
%
$y=0$. All cross sections have been multiplied by the branching ratio of 5.88~\%
for $J/\psi\to\mu^+\mu^-$. As one can see, the
cross sections for PbPb collisions are more than three orders of magnitude larger than
the ones for pPb, and six orders of magnitude larger than those for pp. Indeed, 
the two-photon spectra for heavy ions are enhanced by factors of $Z^2$ for pPb and $Z^4$ for PbPb
up to corrections due to different minimum impact parameters for an ultra-peripheral collisions.
Note also that for instance the cross section for $J/\psi c\bar c$ decreases faster as a function of
$P_T$ for PbPb than for pp. This can be traced back to the harder spectrum of $\gamma$ from p than for Pb.
 
As we have seen above, the expected luminosities for heavy-ion runs are also much
smaller and scale, in fact, like $Z^2$. Taking into account the run duration
($10^7$ sec for pp and $10^6$ sec for pPb and PbPb), the number of events will be
comparable to, if not a bit larger than in the pp runs. Note that ultraperipheral
collisions of lead ions can be triggered on by requiring the detection of a
forward neutron emitted by a lead ion, that was excited during the collision~\cite{Baltz:2007kq}.

In each plot, we show three curves corresponding to the associated production
of $J/\psi$ mesons with photons (dotted) and $c\bar{c}$ pairs (full) directly
through an $S$-wave color-singlet state and with gluons (dashed) through an
intermediate color-octet $S$-wave state. The associated production with photons
and gluons receives contributions from exactly the same Feynman diagrams, so that
the two curves show nearly the same shape. They differ only in normalization
due to different couplings (electromagnetic vs.\ strong) and operator matrix
elements (singlet vs.\ octet), the former difference being slightly $P_T$
dependent due to the running of $\alpha_s$ that we took into account. The
associated production of $J/\psi$ mesons with an additional $c\bar{c}$ pair
allows, however, for different recombinations
of the intermediate charm quarks, leading to a very different and visibly
much flatter $P_T$-distribution than in the other two cases. In particular, this
channel becomes dominant already for $P_T>2$ GeV.

\section{Conclusion}

As we have seen, the mechanism of inclusive production of heavy quark-antiquark
bound states (quarkonia) is still far from being understood. Leading-order
calculations seem to indicate a large contribution from intermediate color-octet
states in agreement with the effective field theory of NRQCD, but this leads to
contradictions with the observed quarkonium polarization. Next-to-leading order
corrections are known to be very important, but are so far only available for a
restricted number of partonic processes.

Proton-proton and heavy-ion collisions at the LHC will soon allow for studies
of very high-energy photon collisions. We have demonstrated that the production of
quarkonia in these collisions will have observable rates at least in pp, but
possibly also in pPb and PbPb scattering. The associated production of $J/\psi$
mesons with photons, gluons and $c\bar{c}$ pairs will lead to experimentally
distinguishable final states,
which are sensitive to different intermediate color states. Together with their
discriminating transverse-momentum spectra, this will hopefully provide crucial
new experimental information on this longstanding issue.

\section*{Acknowledgments}

\noindent We thank P.\ Artoisenet and F.\ Maltoni for sharing their results for
 pp-collisions at 14 TeV with us and D.\ d'Enterria and T.\ Pierzchala for useful
 discussions.

\end{document}